\documentclass[a4paper,11pt]{article}
\pdfoutput=1 

\usepackage{jinstpub} 
\usepackage{siunitx}
\usepackage{xspace}

\newcommand{\apsq}{\texorpdfstring{\ensuremath{\text{Allpix}^2}}{Allpix\textasciicircum 2}\xspace}

\title{\boldmath \apsq~--~Silicon Detector Monte Carlo Simulations for Particle Physics and Beyond}

\author{S. Spannagel}
\author{and P. Sch\"utze}

\affiliation{Deutsches Elektronen-Synchrotron DESY,\\Notkestr. 85, 22607 Hamburg, Germany}

\emailAdd{simon.spannagel@desy.de}

\abstract{\apsq is a versatile, open-source simulation framework for silicon pixel detectors.
Its goal is to ease the implementation of detailed simulations for both single sensors and more complex setups with multiple detectors.
While originally created for silicon detectors in high-energy physics, it is capable of simulating a wide range of detector types for various application scenarios, through its interface to Geant4 to describe the interaction of particles with matter, and the different algorithms for charge transport and digitization.
The simulation chain is arranged with the help of intuitive configuration files and an extensible system of modules, which implement the individual simulation steps.
Detailed electric field maps imported from TCAD simulations can be used to precisely model the drift behavior of the charge carriers, bringing a new level of realism to the Monte Carlo simulation of particle detectors.

Recently, \apsq has seen major improvements to its core framework to take full advantage of multi- and many-core processor architectures for simulating events fully parallel.
Furthermore, new physics models such as charge carrier recombination in silicon have been introduced, further extending the application range.
This contribution provides an overview of the framework and its components, highlighting the versatility and recent developments.}


\proceeding{12$^{\text{th}}$ International Conference on Position Sensitive Detectors\\
12–17 Sep 2021\\
Birmingham}

\begin{document}
\maketitle
\flushbottom

\section{Introduction}
\label{sec:intro}

With the advent of complex silicon detectors such as monolithic sensors or hybrid detectors with 3D sensors, detailed end-to-end Monte Carlo simulations of such devices have become an indispensable tool for detector R\&D.
They are used to optimize the detector design prior to production, to improve the understanding of the signal formation process, or to interpret data of detectors already in operation.
Over the years, many different Monte Carlo codes have been developed, but they were either dedicated to a specific experiment, specialized on a certain silicon detector type, or are in lack of a modular approach that would allow for a wide range of applications.

The \apsq pixel detector simulation framework~\cite{apsq} has been developed as a versatile tool with an emphasis on longevity.
It caters to the diverse needs of the R\&D community by embracing a modular scheme that allows to tailor the simulation pipeline to the individual requirements of the detectors and applications.
This section introduces the guiding development principles and covers some of the recent additions to the framework.

\subsection{Guiding Principles of Development}

In order to ensure sustainable development and flexible software, the following principles have been applied to the development of \apsq.

\paragraph{Integration of Existing Toolkits}

Many powerful tools have been developed and are employed intensively in detector R\&D.
\apsq leverages their capabilities by providing interfaces to integrate them closely into the simulation.

One prominent example is Geant4~\cite{geant4, geant4-2, geant4-3} which is widely used to simulate the interaction of particles with matter.
In \apsq, Geant4 is one of several options to simulate the initial energy deposition in the sensor by ionizing radiation.
Geant4 is an extensive toolkit which allows for the detailed simulation of many interaction processes in arbitrary, user-defined geometries.
However, this complexity is sometimes overwhelming for new users.
The relevant modules in \apsq therefore provide an abstraction layer that auto-generates the geometrical models of the detectors, that translate the key-value configurations of relevant parameters to the respective Geant4 interface commands, and that take care of calling the Geant4 kernel.

Another important tool in silicon detector R\&D are TCAD simulations that solve Poisson's equation in the sensor using detailed doping information.
The resulting field configurations in the sensor allow to draw conclusions on the sensor behavior and to optimize the design.
By providing the possibility of importing static fields from TCAD simulations to complement the Monte Carlo models, \apsq enables time-resolved simulations of the signal formation on a much faster timescale than transient TCAD simulations while in addition including stochastic effects such as Landau fluctuations.

\paragraph{Validation of Algorithms}

Simulations provide insights into physical processes but require detailed validation before being useful for predictions.
The development procedures set in place for \apsq attempt to conduct a validation of new algorithms against either other simulation tools or data before being released.
Validations conducted by the core development team are published as a series of papers~\cite{apsq, allpix-hrcmos, allpix-transient}.
In addition, the continuous integration pipeline of the project repository runs a suite of automated tests for each new version of the framework in order to ensure that the existing algorithms continue to work as expected.

\paragraph{Low Entry Barrier for New Users}

\apsq attempts to facilitate quick starts and first results by providing an extensive documentation in form of a user manual~\cite{apsq_manual}, source code documentation~\cite{apsq-website} and a set of examples for different application scenarios.
The framework is controlled via human-readable configuration files and has support for physical units.
In addition, mailing lists and a forum help connecting users and provide an opportunity to discuss problems.
Since no prior knowledge of programming is required, the framework has already been successfully used in university courses and summer schools on detector instrumentation.

\paragraph{Maintainability of Code}

The development of \apsq follows best practices for software development.
Extensive code reviews are conducted for all contributions via merge requests, and the code base strictly enforces coding conventions as well as consistent formatting of source code.
Static code analysis is performed regularly to detect possible errors and problems at an early stage.
\apsq is published under the permissive open-source MIT license to encourage use in both academia and industry.

\subsection{Recent Developments}

Recently, a new major version -- \apsq 2.0 -- has been released with many changes to the core parts of the software.
Some of the new features will be briefly  discussed in the following while a more exhaustive list can be found in the corresponding release notes~\cite{apsq2-release}.

\paragraph{Multithreading} With the new major version, \apsq supports event-based multithreading.
Events are placed in a central task queue and worker threads pick up new events for processing whenever the previously simulated event is finished.
The number of worker threads can be configured at run-time.
A buffering mechanism is provided in case modules require a deterministic order of events.
In this case, events which are not to be processed by the respective module are cached in the buffer until all previous events have been processed.
The implementation of multithreading in \apsq allows to retain strong reproducibility.
Each simulation conducted with the same configuration and the same seed for the pseudo-random number generators will yield the exact same result, independent of the number of workers chosen or the load and scheduling of the host machine.

\paragraph{Charge Carrier Lifetime \& Recombination}

With the new version it is possible to load doping profiles and to enable doping-dependent lifetime calculations for charge carriers.
This is especially relevant in situations where low electric field regions and high doping concentrations are present, leading to a very fast recombination of charge carriers with the silicon lattice.
Several recombination models have been implemented and can be configured via the configuration file of the simulation.

\paragraph{Charge Carrier Mobility}

Prior to \apsq 2.0, only one charge carrier mobility model was implemented and chosen for all simulations~\cite{jacoboni}.
With the new release, a set of different models with dependencies not only on the electric field, but also on the doping concentration if available, have been introduced and can now be selected via the configuration file.
This is of special interest for sensors with high electric fields or strong doping gradients which affect the mobility and velocity of the charge carriers.
In addition to several models from literature, an \emph{extended Canali} model has been implemented, combining a model with doping concentration dependence~\cite{masetti} and one with a saturation velocity~\cite{canali}.

\section{Application Examples}

Over the past years, \apsq has been used in a variety of different application scenarios.
Many examples have been presented by users at the \emph{2nd Allpix Squared User Workshop}~\cite{apsq-ws}, some of which are briefly summarized in the following.

\subsection{Signal formation in MAPS Prototypes}

\apsq has been used to conduct detailed transient simulations of the signal formation process in monolithic active pixel sensors~\cite{apsqws-cmos}.
Electrostatic fields and weighting potentials from TCAD simulations are used in conjunction with a simulation of the full detection process to obtain realistic time-response distributions for a variety of sensor designs.
Furthermore, \apsq simulations have been used to determine the substrate resistivity of a sensor prototype by comparing the detector performance obtained from simulations with different resistivities to that measured in test beam experiments.
Some of the results of the simulations have been published~\cite{allpix-hrcmos, allpix-transient}.

\subsection{EPICAL-2: Electromagnetic Pixel Calorimeter}

It is foreseen to add a forward calorimeter to the ALICE experiment at the LHC to enhance the experiment's physics reach~\cite{ALICECollaboration:2719928}.
A current technology demonstrator, called \emph{EPICAL-2}, consists of 24 layers of ALPIDE sensors~\cite{MAGER2016434} interleaved with \SI{3}{mm} tungsten absorbers and has been simulated with \apsq~\cite{apsqws-epical}, using the direct interface to Geant4 to simulate the shower development.
Good agreement between data recorded in test beam experiments and the simulation have been found.
The simulation allowed the study of the sensor response to the shower particles, including the longitudinal and transversal shower profiles.
Several adjustments of simulation parameters, such as a more realistic beam profile and energy spectrum, are underway and a publication is in preparation.

\subsection{Dual-Sided Micro-Structured Neutron Detector}

Another application of silicon detector Monte Carlo simulations is the description of charge collection in a dual-sided micro-structured neutron detector~\cite{apsqws-dsmsnd}.
Here, trenches are etched into a silicon sensor and back-filled with LiF which acts as neutron conversion material.
The $\alpha$ and triton secondary particles emerging from the reaction of lithium with thermal neutrons then create charge carriers in the silicon sensor, and \apsq has been used to simulate the charge carrier motion and signal formation.
Some of the simulation results have been published~\cite{dsmsnd}.

\section{Current Developments}

Currently, many features are under development and a new major release is in preparation.
In the following, a few highlights from the current development cycle are presented.

\subsection{Hexagonal Pixel Geometries}

Initially the geometry sub-system of the framework focused on rectangular pixels or strips in a regular matrix pattern.
However, different pixel shapes can be beneficial for certain sensor designs and a more flexible geometry is in preparation.

Most prominently, this will allow the simulation of hexagonal pixels and honeycomb matrices, which are interesting for a number of applications.
Hexagonal pixel shapes avoid problematic field regions in the pixel corners by reducing the maximum distance of the pixel boundary to the center while maintaining the same area.
Furthermore, hexagons provide a symmetry more close to a circle and therefore feature a more uniform sensor response over the pixel area.
The implementation in \apsq allows the usage of different hexagon orientations as well as regular or irregular hexagon shapes with different pitches along the grid axes.

In addition, other geometries making use of the more flexible framework are in preparation, such as radial strip sensors for example used in the ATLAS ITk endcap detectors~\cite{CERN-LHCC-2017-005}.

\subsection{Impact Ionization}

An important effect in the presence of large electric fields in silicon sensors is charge multiplication through impact ionization.
Currently, different ionization models are being implemented and tested in \apsq, and an interface to the \emph{Weightfield2} program~\cite{weightfield} is foreseen.
Similar to the mobility and recombination models, the impact ionization model as well as the multiplication threshold field will be selectable from the main configuration file of the simulation.
Currently, the implementation is undergoing detailed testing and comparison both with other simulation packages and reference data.

\section{Summary \& Outlook}

Silicon Detector Monte Carlo simulations are a vital tool to deepen the understanding and allow the interpretation of the detector performance.
\apsq is a flexible simulation framework for this purpose, which integrates well with existing toolkits, applies validated algorithms and is easy to get started with.
It has a clean and solid code base, provides comprehensive documentation and is used in many areas, of which the simulation of CMOS sensors, electromagnetic calorimeters, or micro-structured neutron detectors are only a few examples.

The framework has seen continuous development and support over several years and a new major version has recently been released, bringing new physics models as well as multithreading capabilities.
Several new features are already underway and will be published in future versions of the framework.

\bibliography{bibliography}

\end{document}